\documentclass[12pt]{article}
\usepackage{amsmath}
\usepackage{amssymb}
\usepackage{mathtext}
\usepackage{graphicx}

\begin{document}
\begin{centering}
\section*{BILLIARDS, INVARIANT MEASURES, AND EQUILIBRIUM
THERMODYNAMICS\footnotemark}
\footnotetext{REGULAR AND CHAOTIC DYNAMICS, V. 5, No. 2, 2000

{\it Received December 12, 1999

AMS MSC 58F36, 82C22, 70F07}}

V.\,V.\,KOZLOV\\
Department of Mechanics and Mathematics\\
Moscow State University\\
Vorob'ievy gory, 119899 Moscow, Russia\\
E-mail: vako@mech.math.msu.su\\
\end{centering}\bigskip

\begin{abstract}
\end{abstract}
The questions of justification of the Gibbs canonical
distribution for systems with elastic impacts are discussed.
A special attention is paid to the description of probability measures with
densities depending on the system energy.

\subsection*{Gibbs Distribution}
Let~$x=(x_1,\,\ldots ,\,x_n)$ be generalized coordinates,
$y=(y_1,\,\ldots ,\,y_n)$ be conjugate canonical momenta of a Hamiltonian
system
with~$n$ degrees of freedom and a stationary Hamiltonian~$H(x,\,y,\,\lambda)$,
where ${\lambda=(\lambda_1,\,\ldots ,\,\lambda_m)}$ are some parameters.
According to Gibbs~[1], a distribution with the probability density
\begin{equation}
\rho=c e^{-\beta H},
\end{equation}
where~$c=const >0,\,\beta =\frac{1}{kT}$ ($T$ is absolute temperature,
$k$ is Boltzmann constant) plays
a key role in statistical consideration of Hamiltonian systems.
The constant~$c$ is chosen due to the normalization condition of density~$\rho$.

Given an invariant measure with the density~(1), we can introduce
an mean energy
\begin{equation}
E(\beta,\,\lambda)=\int H\rho\,d^nx\,d^ny,
\end{equation}
and average generalized forces (constraint reactions~$\lambda =const$),
corresponding to the parameters~$\lambda$:
\begin{equation}
\Lambda_i=-\int {\frac{\partial{H}}{\partial\lambda_i}} \rho\,d^nx\,d^ny\,,\quad 1\le i\le m\,.
\end{equation}

Relations~$L_i=f_i(\beta,\,\lambda)$ are considered as equations of state.

As it was shown by Gibbs, 1-form of heat gain
\begin{equation}
\omega=dE+\sum_1^m \Lambda_i\,d\lambda_i
\end{equation}
satisfies the axioms of thermodynamics: the form~$\beta\omega$ is exact ($\beta\omega =dS$,
where~$S(\beta,\,\lambda)$ is the entropy of a thermodynamical system). In particular,
the form~$\omega$ is an exact 1-form under fixed values of~$\beta$. Thus, according
to Gibbs, to any Hamiltonian system (provided that the integrals~(2) and~(3)
exist and depend smoothly on~$\lambda $ and~$\beta$) there can be associated a
thermodynamic system with external parameters~$\lambda_1,\,\ldots
,\,\lambda_m$, the internal energy~(2), and the equations of state~(3).
The relations~(2) and~(3) can be simplified by introducing statistical
integral
\begin{equation}
Z(\beta,\,\lambda)=\int e^{-\beta H}\,d^nx\,d^ny\,.
\end{equation}

Hence,
\begin{equation}
E=-{\frac{\partial{\ln Z}}{\partial \beta}},\quad \Lambda_i=\frac
{1}{\beta}{\frac{\partial{\ln Z}}{\partial\lambda_i}}
\end{equation}

and therefore~$\beta\omega =dS$, where
\begin{equation}
S=\ln Z-\beta{\frac{\partial{\ln Z}}{\partial\beta}}\,.
\end{equation}

\subsection*{Thermodynamics of Billiards}

Billiard is a mass particle performing the inertial motion in domain~$D$ of
three-dimensional Euclidean space and reflecting
elastically from its boundary~$\partial D$. We can consider a more general
case when there are~$n$-identical particles in the domain~$D$ not
interacting with each other (in particular, not colliding with each
other). Such a system is a universally recognized model of rarefied
perfect gas.

Let~$q_i =(x_i,\,y_i,\,z_i)$ be a set of Cartesian coordinates of the
i-th particle of unit mass with
momentum~$p_i = (\dot{x}_i,\,\dot{y}_i,\,\dot{z}_i)$. Dynamics of the system
in the domain~$D$ is defined by a Hamiltonian
$$
  H=\sum \frac{p_i^2}{2}\,.
$$

Since this function does not contain any information about the
geometry of the domain~$D$, equations~(2) and~(3) are not applicable
immediately. In this case, one can apply the following procedure:
statistical integral~(5) is calculated
first, and then
relations~(6) are used. In our case
$$
Z=\int_{\mathbb{R}^{3n}}\int_{D^n}e^{-\beta H}\,d^3 p_1\ldots d^3 p_n\,
d^3 q_1\ldots d^3 q_n=\Bigl(\frac{2\pi}{\beta}\Bigr)^{\frac{3n}{2}} v^n\,,
  \eqno{(8)}
$$
where~$v$ is the volume of~$D$. Therefore,
the only external parameter~$\lambda$ is the volume~$v$;
a conjugate variable~$\Lambda$ is the gas
pressure~$p$ inside~$D$. Taking into account~(8), from~(6) we obtain
known equations of a perfect gas
$$
E=\frac {3}{2} kT\,,\quad p=\frac{nkT}{v}\,.
  \eqno{(9)}
$$

Billiards, being systems with one-way constraints, are idealization
of ordinary mechanical systems with smooth Hamiltonians. When a particle
hits the wall, the wall deforms giving rise to great elastic forces
which push the particle back into D. These elastic forces are modeled
by potential~$V_{\nu}(q)$. It equals zero in~$D$
and~$\frac{\nu f^2(q)}{2}$ outside~$D$. Here~$f$ is a smooth function
that defines the boundary equation~$\partial D\colon f(q) = 0$. The large
constant~$\nu$ plays a role of elasticity coefficient. It is assumed
that the boundary does not contain critical points of the function~$f$; in
particular, boundary~$\partial D$ is a smooth regular surface. As was shown
in~[2], as~$\nu\to\infty$, solutions of a system with the Hamiltonian
$$
H=\frac{p^2}{2}+V_\nu(q)
  \eqno{(10)}
$$
tend to the motions of a system with elastic reflections in~$D$.

Application of the Hamiltonian~(10) gives corrections to the
expression of statistical integral which depend on the area~$\sigma$
of the boundary of~$D$. Thus, the area~$\sigma$ should be as an
external parameter of the perfect gas as a thermodynamic system; pressure
will be the function of not only volume and temperature, but also of
the surface area of a vessel.

The meaning of the correction is that the volume~$v$ in~(8)
is replaced by
$$
v+\sqrt{\frac{\pi}{2\nu\beta}}\sigma+O\bigl(\nu^{-\frac{3}{2}}\bigr)
  \eqno{(11)}
$$
provided that~$f$ does not have critical points outside~$D$.
Taking this fact into account, the equation of
internal energy~$E$ remains the same and the state equation~(9) is replaced by


$$
p=\frac{nkT}{v+\sqrt{\varkappa T}\sigma}\,,\quad \varkappa=\frac{\pi k}{\nu}\,.
  \eqno{(12)}
$$

Since~$\sigma$ is a new thermodinamical parameter, we should introduce a
conjugate variable
$$
\eta=\frac {1}{\beta}{\frac{\partial{\ln Z}}{\partial\sigma}}=\frac{nkT\sqrt{\varkappa T}}
{v+\sqrt{\varkappa T}\sigma}\,.
  \eqno{(13)}
$$

The relations~(12) and~(13) constitute a total system of the state equations.

Let us indicate the deduction of the formula~(11). To do so, we use an obvious
formula
$$
  \int_{\mathbb{R}^3} e^{-\beta V_\nu}\,d^3q=v+\int_{f\ge 0}e^{-\frac{\beta\nu f^2}{2}}d^3q\,.
$$
According to the saddle-point method, the basic contribution to the
asymptotics of the second integral as~$\nu\to\infty$ is made by the
critical points of the potential~$V$. In accordance with the
assumption, $df\ne 0$ for~$f>0$. Consiquently, a set of
critical points coincides with the boundary~$\partial D=\{f=0\}$. A
non-isolation of the critical points results in a certain difficulty
under usual application of the saddle-point method. Let us pass
(locally) into a neighdourhood of the boundary to semigeodesical
coordinates~$u_1,\,u_2,\,u_3$, where~$f\equiv u_1$~[3]. In these
variables, the Euclidean metric is written in the form
$$
du_1^2+a\,du_2^2+2b\,du_2\,du_3+c\,du_3^2\,,
$$
where~$a,\,b,\,c$ are  smooth functions of~$u_s$. In these
variables the desired integral is replaced asymptotically by the
integral
$$
\int_{u_1\ge 0} g(u_1)e^{-\frac{\beta\nu u_1^2}{2}}du_1\,,
  \eqno{(14)}
$$
where
$$
g=\iint\limits_{\partial D}\sqrt{G}\,du_2\,du_3\,,\quad G=ac-b^2>0\,.
$$

Then, with the help of a standard method~[4], we obtain the asymptotics of the
integral~(14):
$$
g(0)\sqrt{\frac{\pi}{2\nu\beta}}+O \bigl(\nu^{-\frac {3}{2}}\bigr)\,.
$$

Note now that~$g(0)=\sigma$.

\subsection*{Probability Distribution}
Now a rigorous deduction of the Gibbs
distribution is given only for the case of vanishing interaction of
individual subsystems. A classical Darwin-Fauler approach represents an
asymptotical (as~$n\to\infty$) deduction of Gibbs distribution from
the general principles of dynamics in the assumption of the ergodic
hypothesis. As it is observed by A.\,Ya.\,Hinchin~[5], this approach
repeats in fact the previous mathematical results, connected with the
limiting probability theorems.

In~[6] there suggested another deduction of distribution~(1).
It is based on the fact that the probability density is a
single-valued first integral~[1]. With the help of Poincar\'e method,
the conditions, under which motion equations of interacting subsystems
do not admit integrals of~$C^2$-class, independent of energy
integral, are indicated. These conditions are constructive and,
obviously, less strong than the assumption of ergodicity. Moreover,
 a natural Gibbs postulate about thermodynamical equilibrum of
subsystems under vanishing interaction is used in [6].

A statistical analogue of this argument is the deduction of a normal
distribution, suggested by Gauss. He does not use the central limiting
theorem, but the postulate that a sample mean is an estimate of
maximum of probability at the finite number of observations $n\ge
3$ (see [7], [8]).

In connection with the above-said it is usefull to set in order the
hierarchy of Hamiltonian dynamical systems with respect to the degree
of their arbitrariness. Let us fix the phase space~$\mathbb{P}$ of
dimention~$2n\ge 4$ with analitical structure and introduce into
consideration a set~$\mathcal{H}$ of all Hamiltonian systems on~$\mathbb{P}$
with analytical Hamiltonians. Certainly, it is supposed that the
property of Hamiltonians to be analitical on~$\mathbb{P}$ is concordant with
analytical structure of~$\mathbb{P}$ itself.

We introduce a sequence of embedded into each other sets of~$\mathcal {H}$:
$$
\mathcal M\subset \mathcal E\subset \mathcal T\subset \mathcal K^0\subset
\mathcal K^1\subset \ldots\subset \mathcal K^\infty\subset \mathcal A\,.
  \eqno{(15)}
$$
Here, $\mathcal M,\,\mathcal E$ and~$\mathcal T$ are the set of systems, which
respectively possess the properties of intermixing, ergodicity and
trasitivity on the energy $(2n-1)$-dimentional
surfaces. Further, $\mathcal K^s$ is the set of systems, which do not
admit the first integrals of  smoothness class of~$C^s(\mathbb{P})$ not
depending on the energy integral. In addition, the case~$s=0$
corresponds to the continuous integrals: they are locally unstable on
the surfaces of the level of energy integral and take equal values on
the trajectories of the Hamiltonian system. The symbol~$\mathcal A$  denotes
Hamiltonian systems, which do not admit an additional analytical
integral.

One can deduce an analogous chain of embedded sets  for the systems
with elastic reflection as well.

First of all we should make sure that the neighbouring sets in the
chain~(15) do not coincide with each other. The inequalities~$\mathcal
M\ne \mathcal E$, $\mathcal E\ne \mathcal T$, $\mathcal T\ne \mathcal K^o$
can be much easily demonstrated by the examples of area preserving
mappings of~$T$ of the two-dimentional torus~$\mathbb{T}^2=\{x,\,y\mod 2\pi\}$.
Such mappings can be treated as Poincar\'e mappings of the energy
manifolds cuts of Hamiltonian systems with two degrees of freedom. A
classical example of mixing transformation drives an automorphism of a
torus, given by a uni-modular matrix
$$
  \left[
  \begin{aligned}
  2 \mbox{ }& 1 \\
  1 \mbox{ }& 1\\
  \end{aligned}
  \right].
$$

The shifts~$x\to x+a$, $y\to y+b$, where numbers~$a,\,b$ and~$2\pi$
are rationally incommensurable, provide us with known examples of
ergodic, but not mixing, transformations. Thus,~$\mathcal M\ne \mathcal E$. It
is considerably more difficult to give examples of transitive, but not
ergodic, transformations with an invariant measure. For the first
instances of such transformations we cite the work by
L.\,G.\,Shnirelman~(1930) and A.\,Bezikovich~(1937). They  considered continuous
automorphisms of a circle. Smooth modifications of such
transformations are indicated in~[9].


To proof the inequality~$\mathcal T\ne\mathcal K^o$ we use an example of
transitive area preserving transformation~$T$ of the square~$K^2$
which leaves the points on its boundary immovable. Such an example is built by
Oxtoby~[10] with the help of
theory of categories of sets. Let us take four such squares and
form one square of quadruplicated area out of the four (see Figure~1).
Identifying opposite sides, we will obtain two-dimentional
torus, where the mapping~$T\colon K^2\to K^2$ is naturally prolonged
to the contituous area-preserving mapping~$T\colon\mathbb {T}^2\to\mathbb {T}^2$. We
need hardly mention that this transformation will no longer be
transitive. Still it does not admit non-constant continuous integrals.
It would be interesting to provide an analytical example of the
transformation from the set of~$\mathcal K^o\setminus\mathcal T$.

Inequalities~$\mathcal K^k\ne\mathcal K^{k+1}(k=0,\,1,\,\ldots ,\,\infty)$
and~$\mathcal K^{\infty}\ne\mathcal A$ are derived from the results of~[11] (see
also~[12]), where the examples of analytical Hamiltonian systems, not
possessing additional integrals of $C^k(C^{\infty})$--class, but at
same time not admitting integrals of~$C^{k+1}\bigl(C^{\omega}\bigr)$--class, are
indicated.

Let us consider one of the links of the chain of inclusions of~(15),
say, $\mathcal T\subset\mathcal K^o$. The question is, which of the two sets is
more massive: $\mathcal T$ or~$\mathcal K^o\setminus\mathcal T$. Apparently, the
second. However, the answer to this question (as well as its
formulation) depends on the introduced topology in the space~$\mathcal K^o$.
Analogous assumptions are probably valid for any pair of the
neighbouring sets in~(15).

Classes of systems from~(15) may be laid out into a wider class of
systems, which do not admit additional single-valued complex-analytic
first integrals. An obstacle to the existence of single-valued
holomorphic integrals is the branching of the solutions of Hamiltonian
systems in the plane of complex time. The discussion of this range of
questions one can find in the work~[13].

If we remain within the real examination, then the class~$\mathcal A$ admits
a natural extension for the dynamics of natural machanical systems.
They are decribed by the Hamiltonians of the form~$H=T+V$, where~$T$
is a kinetic energy, a positively defined quadratic form with respect
to the momenta, and~$V$ is a potential energy, a function on the
configuration space.  All known integrals of such systems are
polynomials in  momenta with single-valued,
coefficients on configuration space,
(or functions of such polynomials). In analytical
case, these coefficients are also represented by analytical functions.
We can show that the existence of an additional polynomial integral of
the system with the Hamiltonian~$H=T+V$ is equivalent to the existence
of an integral of the system with the Hamiltonian~$H=T+\varepsilon V$
($\varepsilon $ is a small parameter) as the series in terms of powers of
$\varepsilon$.

This problem is more simple and since Poincar\'e times
there have been proposed
 efficient methods for its solution~[13].
The existence conditions of additional polynomial integral of a
plane billiard are obtained with the help of complex variable function~[14].

The issue of whether a certain Hamiltonian system
belongs to the class~$\mathcal A$ is more complex. But essential
advances have been made in this field as well, especially for the case
with small number of degrees of freedom~[13]. Difficulties become
much more severe as we move towards the beginning of the chain~(15).
Thus, according to Kac~[15], an efficient verification of the ergodic
property of a dynamical system is a nearly hopeless problem. Moreover,
in many important cases, from the application viewpoint, ergodic hypothesis
is refuted by the results of KAM theory. For instance, as it was
established by Lazutkin~[16], a billiard inside a plane convex curve
(of~$C^2$--class of smoothness) is not ergodic. It does not even possess
the transitive property. Lack of ergodicity in spatial case was proved
 in~[17] under some additional conditions.
These examples are directly related with the deduction of Gibbs distribution
for the perfect gas.

For small perturbation of an integrable Hamiltonian system with two
degrees of freedom, Kolmogorov tori cut a three-dimentional
energy surfaces. Therefore, a perturbed system can no longer be
transitive. On the other hand, as it was noted by Arnold, such systems
admit a nonconstant continuous integral that takes constant values
in slits between Kolmogorov tori. It's not quite clear yet whether
such systems have locally nonconstant
continuous first integrals which are not identically constant in any
neighbourhood of every point of the energy surface. A simpler problem is
whether perturbed systems of general kind with two degrees of
freedom admit nonconstant integrals of $C^1$--smoothness class.

For systems with~$n\ge 3$ degrees of freedom, the slits between
Kolmogorov tori form a connected set everywhere densely filling a
five--dimensional energy manifolds. Therefore, a principal possibility
of the appearance of transitive property arises. This is one of the
exact statements of the known hypothesis of diffusion in perturbated
multidimensional Hamiltonian systems. For the purpose of statistical
mechanics this diffusion hypothesis can be formulated in a less
restricted fashion: is it true that under great~$n$ a perturbed
Hamiltonian system of general form does not admit nonconstant
continuous (or even smooth, of~$C^1$--class) first integrals
on~($2n-1$)-dimensional energy surfaces? In fact, it is sufficient that
this property appeared under a small fixed value of perturbing
parameter~$\varepsilon$ and a great value of~$n$ of weakly interacting
subsystems.

\subsection*{Generalized entropy}

Our observations described in previous Section result in a natural assumption
that the density of probability distribution~$\rho$ is a function of~$H$. The question is: what
makes Gibbs distribution different from all other distributions of this
kind?

Let~$z\to f(z)$ be a nonnegative real function of one variable, $f'$ be its
derivative. Following Gibbs, we will consider probability density
$$
\rho=\frac{f(\beta H)}{\int f(\beta H)\,dx\,dy}
  \eqno{(16)}
$$
assuming that the integral converges over the whole phase space. Here
again~$\beta^{-1}=kT$. When~$f=ce^{-z},\,c=const\ne 0$, we shall obtain
Gibbs distribution. We could consider a more general case, when the
function~$f$ depends also on
external parameters~$\lambda$ (as well as the Hamiltonian~$H$). But we shall not
follow this case.

Let us calculate an average energy~$E$ and generalized forces~$\Lambda$ using~(2)
and~(3), with density~$\rho$ determined by~(16). Then we can compose
1-form of heat gain in accordance with~(4). Using direct calculations
we can prove

{\bf Theorem.} {\it
The form~$\omega$ satisfies axioms of thermodynamics if}
$$
\int \frac{\partial{H}}{{\partial\lambda_i}}f\,dx\,dy\int \frac{\partial{H}}{{\partial\lambda_j}}f'\,dx\,dy=
\int\frac{\partial{H}}{{\partial\lambda_j}}f\,dx\,dy\int \frac{\partial{H}}{{\partial\lambda_i}}f'\,dx\,dy\,.
  \eqno{(17)}
$$
{\it for all~$1\le i,\,j\le m$ and}
$$
\int Hf\,dx\,dy\,\int \frac{\partial{H}}{{\partial\lambda_i}}f'\,dx\,dy=
\int \frac{\partial{H}}{{\partial\lambda_i}}f\,dx\,dy\, \int Hf'\,dx\,dy
  \eqno{(18)}
$$
{\it for all~$1\le i\le m$.}

It is obvious that for the function~$f(z)=ce^{-z}$ these conditions are met.
Equalities~(17) and~(18) can be rewritten as follows
$$
\Lambda_i\frac{\partial{F}}{{\partial\lambda_j}}=\Lambda_j\frac{\partial{F}}{{\partial\lambda_i}}\,,\quad
(1\le i,\,j\le m)\,,
  \eqno{(19)}
$$
$$
\frac{E}{\beta}\frac{\partial{F}}{\partial\lambda_i}=
-\Lambda_i\frac{\partial{F}}{{\partial\beta}}\,,\quad (1\le i\le m)\,,
  \eqno{(20)}
$$
where
$$
  F=\int f(\beta H)d^n x\,d^n y\,.
$$
By analogy with Gibbs case, the function~$F$ can be called a generalized
statistical integral.\goodbreak

From~(19) and~(20) follows the existence of function~$x(\beta
,\,\lambda_1,\,\ldots ,\,\lambda_m)$, such that
$$
\Lambda_i=-\frac{\varkappa}{\beta}\frac{\partial{F}}{\partial\lambda_i}\,,\quad
E=\varkappa\frac{\partial F}{\partial\beta}\,.
  \eqno{(21)}
$$
Therefore, the form of heat gain takes the form
$$
\omega=d\Bigl(\varkappa\frac{\partial
F}{\partial\beta}\Bigr)-\sum\frac{\varkappa}{\beta}\frac{\partial{F}}{\partial\lambda_i}\,
d\lambda_i\,.
$$
Axioms of thermodynamics impose constraints on the form of
function~$x$. From~(19) we obtain a series of inequalities
$$
  \frac{\partial{\varkappa}}{\partial\lambda_i}\frac{\partial{F}}{\partial\lambda_j}-
  \frac{\partial{\varkappa}}{\partial\lambda_j}\frac{\partial{F}}{\partial\lambda_i}=0\,,\quad
  (1\le i,\,j\le m)\,, \eqno{(22)}
$$
and the equation~(20) yields relations
$$
  \frac{\partial{\varkappa}}{\partial\beta}\frac{\partial{F}}{\partial\lambda_j}-
  \frac{\partial{\varkappa}}{\partial\lambda_i}\frac{\partial{F}}{\partial\beta}=0\,,\quad (1\le i\le m)\,.
  \eqno{(23)}
$$

Equalities~(22) and~(23) denote that functions~$\varkappa$ and~$F$ are dependent.
Therefore, we can write that~$\varkappa=\varkappa(F)$, at least locally.

Let~$\Phi$ be antiderivative of~$\varkappa(\cdot)$. Then equalities~(21)
take a simpler form
$$
\Lambda_i=-\frac{1}{\beta}\frac{\partial{\Phi}}{\partial\lambda_i}\,,\quad
E=\frac{\partial{\Phi}}{\partial\beta}\,.
  \eqno{(24)}
$$
Hence
$$\begin{aligned}
\beta\omega & = \beta\,d\Bigl(\frac{\partial{\Phi}}{\partial\beta}\Bigr)-
\sum \frac{\partial{\Phi}}{\partial\lambda_i}\,d\lambda_i=\\
& =
d\Bigl(\beta\frac{\partial{\Phi}}{\partial\beta}\Bigr)-\frac{\partial{\Phi}}{\partial\beta}\,d\beta-
\sum \frac{\partial{\Phi}}{\partial\lambda_i}\,d\lambda_i=\\
& = d\Bigl(\beta\frac{\partial{\Phi}}{\partial\beta}-\Phi\Bigr)\,.
\end{aligned}$$
The function
$$
S=\beta\frac{\partial{\Phi}}{\partial\beta}-\Phi
  \eqno{(25)}
$$
is called an entropy in thermodynamics.

The form of this function suggests that Legendre transform over~$\beta$
should be applied. Assuming that
$$
  \frac{{\partial}^2\Phi}{\partial\beta^2}\ne 0\,,
$$
from the second relation of~(24) we will obtain~$\beta$ as a function of~$E$
and~$\lambda$. We will assume~$E,\,\lambda_1,\,\ldots ,\,\lambda_m$ independent variables.
Then~$ S = S(E, \lambda)$ and from~(25) we will obtain potential form of
basic thermodynamic relations~(24):
$$
  \beta=\frac{\partial S}{\partial E},\quad \beta\Lambda_i=\frac{\partial{S}}{\partial\lambda_i}\quad (1\le i\le m).
$$

\subsection*{The Perfect Gas}

Let us apply relations from previous Section to the perfect gas inside
domain~$D$ of the three-dimentional Euclidean space; let~$v$ be the
volume of~$D$. Remembering that the perfect gas is a totality of~$n$
equal and not interacting particles
performing the inertial motion inside~$D$ and reflecting
elastically from its boundary~$\partial D$. When taking into account
arbitrarily small interaction of particles, we will obtain a system
without additional integrals and therefore we can consider that
the density of probability distribution
is a function of total energy. Let particle interaction tends
to zero; then we will obtain simple equations for average energy and
state equations; these equations define thermodynamics of a
simplified system, i.~e. the perfect gas. Let particle mass be equal to
unit. Hence, the  Hamiltonian for the perfect gas will be determined by the
following equation
$$
  H=\sum\frac{p_i^2}{2}\,,
$$
where~$p_i = (\dot x_i,\dot y_i,\dot z_i)$ is momentum of the $i$-th particle;
let~$q_i = (x_i,\,y_i,\,z_i)$ be  its Cartesian coordinates.

The formula for internal energy has the form
$$
  E=\frac{\int_{\mathbb{R}^{3n}}\int_{D^n} \frac {1}{2} \sum p_i^2 f\Bigl(\frac{\beta}{ 2}\sum
p_i^2\Bigr)\,d^{3n}p\,d^{3n}q}
{\int_{\mathbb{R}^{3n}}\int_{D^n} f\Bigl(\frac{\beta} {2}\sum
p_i^2\Bigr)\,d^{3n}p\,d^{3n}q}\,.
$$
It is  independent of volume~$v$:
$$
E(\beta)=\frac{a}{b\beta}\,,
  \eqno{(26)}
$$
where
$$\begin{aligned}
a & = \int_{\mathbb{R}^{3n}} \frac{1}{2} \sum u_i^2 f\Bigl(\frac {1}{2}\sum
u_i^2\Bigr)\,d^{3n}u\,,\\
b & = \int_{\mathbb{R}^{3n}}f\Bigl(\frac {1}{2}\sum
u_i^2\Bigr)\,d^{3n}u\,.
\end{aligned}$$
Variables~$p$ and~$u$ are connected by simple relations: $u_i=\sqrt{\beta}p_i$.

Assuming for simplicity~$3n = m + 2$, we will pass from~$u_1,\,\ldots ,\,u_{m+2}$
to spherical coordinates~$r,\,\theta_1,\,\dots ,\,\theta_m,\,\varphi$
using the following equations
$$\begin{aligned}
 u_1 & = r\cos\theta_1\,,\\
 u_2 & = r\sin\theta_1\cos\theta_2\,,\\
 u_3 & = r\sin\theta_1\sin\theta_2\cos\theta_3\,,\\
\ldots & \ldots\ldots\ldots\ldots\ldots\ldots\ldots\ldots\\
 u_m & = r\sin\theta_1\sin\theta_2\ldots \sin\theta_{m-1}\cos\theta_m\,,\\
 u_{m+1} & = r\sin\theta_1\sin\theta_2\ldots\sin\theta_m\cos\varphi\,,\\
 u_{m+2} & = r\sin\theta_1\sin\theta_2\ldots\sin\theta_m\sin\varphi\,.
\end{aligned}$$

Here~$r\ge 0,\, 0\le \theta_j\le\pi\,(1\le j\le m)$ and~$\varphi
\mod 2\pi$ is an
angular coordinate.

In the new coordinates
$$
\begin{aligned}
b={}&\int_0^\infty r^{m+1}f\Bigl(\frac{r^2}{2}\Bigr)\,dr
\int_0^\pi\dots\int_0^\pi\int_0^{2\pi}(\sin\theta_1)^m(\sin\theta_2)^{m-1}
\ldots(\sin\theta_m)\,d\theta_1\dots d\theta_m\,d\varphi=\\
={}&\frac{2\pi^{1+\frac{m}{2}}}{\Gamma\Bigl(1+\frac {m}{2}\Bigr)}
\int_0^\infty r^{m+1}f\Bigl(\frac{r^2}{2}\Bigr)\,dr\,,
\end{aligned}
  \eqno{(27)}
$$
where~$\Gamma$ is Euler's gamma-function. By analogy,
$$
a=\frac{2\pi^{1+\frac{m}{2}}}{\Gamma\Bigl(1+\frac {m}{2}\Bigr)}
\int_0^\infty \frac{r^{m+3}}{2}f\Bigl(\frac{r^2}{2}\Bigr)\,dr\,.
  \eqno{(28)}
$$

Now we calculate generalized statistical integral:
$$
F=\int_{\mathbb{R}^{3n}}\int_{D^n} f\Bigl(\frac{\beta}{ 2}\sum p_i^2\Bigr)\,d^{3n}p\,
d^{3n}q=\frac{bv^n}{(\sqrt{\beta})^{3n}}\,.
  \eqno{(29)}
$$
According to~(21)
$$
  E=\varkappa \frac{\partial{F}}{\partial\beta}\,.
$$
Therefore, taking into account~(26) and~(29),
$$
  \varkappa=-\frac{2a(\sqrt{\beta})^{3n}}{3nb^2v^n}\,.
$$

Applying the first equation~(21), we obtain state equations
$$
  \Lambda=-\frac{\varkappa}{\beta}\frac{\partial {F}}{\partial v}=\frac{2a}{3bv\beta}\,.
$$
Denoting pressure~$\Lambda$ by~$p$ in accordance with established thermodynamical
notation, we arrive at a more usual form of
state equation:
$$
  pv=\frac{ 2}{3} \frac {a}{b} kT\,.
  \eqno{(30)}
$$

Now let~$f(z) = e^{-z}$. Thus,
$$
  b=\frac{1}{m+2}\int_0^\infty e^{-\frac{r^2}{2}}\,dr^{m+2}=\frac{2a}{m+2}\,.
$$
Hence, $\frac{a}{b}=\frac{(m+2)}{2}=\frac{3n}{2}$ and state equation~(30)
trnasforms into the classical Clapeyron equation:
$$
  pv=nkT\,.
  \eqno{(31)}
$$

Now assuming that state equations~(30) and~(31) are identical under
any~$n$,  we can ask the following question. Is it true that frequency
function will be of Gibbs form, i.\,e. $f(z) = \exp(-z)$? The answer
appears to be negative. Actually, (30) and~(31) are identical if
$$
  \frac {a}{b}=\frac{3n}{2}=\frac{m+2}{2}\,,\quad m=3n-2=1,\,4,\,7,\,\ldots\,.
$$
With account of~(27) and~(28) these equations take the following form
$$
\int_0^\infty r^{m+3}f\Bigl(\frac{r^2}{2}\Bigr)dr=(m+2)
\int_0^\infty r^{m+1}f\Bigl(\frac{r^2}{2}\Bigr)dr\,.
  \eqno{(32)}
$$
Let~$f$ be decreasing at infinity faster than any exponential function.
Then by part-wise integrating we can represent~(32) as
follows
$$
\int_0^\infty \Bigl[f'\Bigl(\frac{r^2}{2}\Bigr)+f\Bigl(\frac{r^2}{2}\Bigr)\Bigr]
r^{m+3}\,dr=0
  \eqno{(33)}
$$
for all~$m+3=3n+1=4,\,7,\,10,\,\ldots$. If this equality was true for
all non-negative~$m + 3$, then according to classical momenta theory~[18],
the expression in the square brackets of~(33) would be equal to zero.
Hence~$f'+f=0$ and, therefore, $f = ce^{-z},\,c = const$. However, (33)
is not valid for the ``majority'' of
integer values of~$m + 3$. Hence, it follows that there is an
infinite-dimensional space of frequency functions dependent on total
energy only, which result in classical thermodynamical relations for
the perfect gas.

The work has been partially supported by RFBR (grant No. 99-01-0196) and INTAS (grant No. 96-0799) foundations.

\begin{figure}[ht!]
$$
\includegraphics{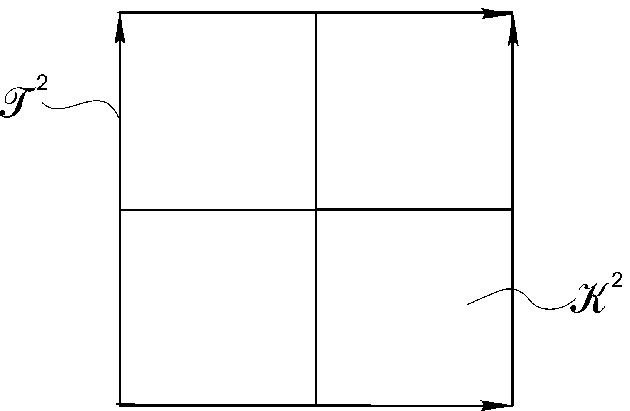}
$$
\caption{}
\end{figure}


\begin{thebibliography}{99}
  \bibitem{Gibbs}
{\it G.\,V.\,Gibbs.}
Thermodynamics. Statistical Mechanics. Moscow, Nauka, 1982,
P.\,384.

  \bibitem{Kozlov1}
{\it V.\,V.\,Kozlov.}
A Constructive Method of Justification of the  Theory
of Systems with Unilateral Constraints. Prikl. Mekh. i Mat.,
1988, V.\,52, Iss.\,6, P.\,883-894.

\bibitem{Rashev}
{\it P.\,K.\,Rashevsky.}
Riemannian Geometry and Tensor Analysis. Moscow, Nauka,
1967, P.\,664.

\bibitem{Brein}
{\it N.\,G.\,Brain.}
Asymptotic Methods in Analysis. Moscow, IL, 1961, P.\,247.

\bibitem{Hinchin}
{\it A.\,Ya\,Hinchin.}
Mathematical Grounds for Statistical Mechanics.
Moscow--Leningrad, Gostekhizdat, 1943, P.\,128.

\bibitem{Kozlov2}
{\it V.\,V.\,Kozlov.}
       Canonical Gibbs Distribution and Thermodynamics of Mechanical Systems
       with a Finite Number of Degree of Freedom.
       Regular and Chaotic Dynamics, 1999, V.\,4, No. ,2, P.\,44--54.

  \bibitem{Wittek}
{\it E.\,T.\,Whittaker, G.\,Robinson.}
The Calculus of Observations. Blackil and Son, 1928.

  \bibitem{Kachan}
{\it A.\,M.\, Kachan, Yu.\,V.\,Linnik, S.\,R.\,Rao.}
Characterizational Problems of
Mathematical Statistics. Moscow, Nauka, 1972, P.\,656.

\bibitem{Sidorov}
{\it E.\,A.\,Sidorov.}
Smooth Topological Transitive Systems. Mathematical
Notes, 1968, V.\,4, No. 6, P.\,751--759.

\bibitem{Oxtoby}
{\it J.\,C.\,Oxtoby.}
        Note of Transitive Transformations.
        {\it Proc. Mat. Acad. Sci. U.S.}, 1937, V.\,23, P.\,443--446.
  \bibitem{Kozlov3}
{\it  V.\,V.\,Kozlov.}
        Phenomena of Nonintegrability in Hamiltonian Systems.
        {Proc. Int. Congr. Math. Berkeley. California.} USA,
        1987, P.\,1161--1170.
  \bibitem{Moshevitin}
{\it N.\,G.\,Moshchevitin.}
On Existence and Smoothness of an Integral of a
Hamiltonian System with a Defined Form. Mathematical Notes, 1991, V.\,49,
No. 5, \mbox{P.\,80--85.}

\bibitem{Kozlov4}
{\it V.\,V.\,Kozlov.}
        Symmetries, Topology and Resonances in Hamiltonian Mechanics.
        Springer--Verlag, 1996, P.\,378.

\bibitem{Bolotin}
{\it S.\,V.\,Bolotin.}
Birkhoff Integrable Billiards. Vestnik MGU, Ser. Mat.,
Mekh., 1990, No. 2, P.\,33--36.

  \bibitem{Kac}
{\it M.\,Kac.}
        Probability and Related Topics in Physical Sciences.
        Intersience Publishers, 1957.

  \bibitem{Lazut}
{\it V.\,F.\,Lazutkin.}
A Convex Billiard and Eigenfunctions of the Laplace Operator.
Leningrad, LGU Publishers, 1981, P.\,196.

\bibitem{Sva}
{\it N.\,V.\,Svanidze.}
Existence of Invariant Tori for a Three-Dimentional
Billiard Located in a Neighbourhood of a ``Closed Geodesic on the Domain
Boundary''. UMN, 1978, V.\,33, Iss.\,4, P.\,225--226

  \bibitem{Ahiezer}
{\it N.\,I.\,Ahieser.}
Classical Problem of Moments. Moscow, Nauka, 1961,
P.\,310.
\end{thebibliography}
\end{document}